# Tangential component of the YORP effect

Oleksiy Golubov[1], Yurij N. Krugly[2]

[1]*Astronomisches Rechen-Institut, ZAH, University of Heidelberg, Mönchhofstraße 12-14, Heidelberg 69120, Germany*

[2]*Institute of Astronomy of Kharkiv National University, Sumska Str. 35, Kharkiv 61022, Ukraine*

Version: 2012 February 26

Proposed running head: Tangential YORP

Editorial correspondence to:

Oleksiy Golubov

Astronomisches Rechen-Institut

Moenchhofstr. 12-14

69120 Heidelberg

Germany

Phone: 0049-6221-54-1825

Fax: 0049-6221-54-1888

E-mail address: golubov@ari.uni-heidelberg.de

**Abstract**

This article discusses how re-emission of absorbed solar light by centimeter- to decimeter-sized structures on the surface of an asteroid can create a component of the recoil force parallel to the surface. Under certain conditions the west side of stones appear to be on average slightly warmer than their east sides, thus experiencing a stronger recoil force and increasing the rotation rate of the asteroid. We study this effect, called the Tangential YORP effect, in a toy model, replacing stones with walls and simulating heat conductivity in them. We discuss general trends of the effect, estimate its magnitude, and find it to be comparable to the normal YORP effect determined by gross-scale asymmetry of the asteroid. The existence of this effect would modify the predictions of the YORP acceleration of asteroids.

# 1. Introduction

The YORP effect is a torque created by recoil forces of light scattered and re-emitted by the surface of an asteroid (Rubincam, 2000; Bottke et al., 2006). In simulations of YORP, shape models of limited resolution are used. Normally the smallest surface features resolved by models delivered by space missions have sizes of order of meters. The smallest structures ordinarily resolved by radar models reach only tens of meters. Models obtained by the light curve inversion method usually resolve only structures spanning over hundreds of meters. Even synthetic shape models used for YORP simulations can't have too high resolution because of limited computation capabilities. Statler (2009) discussed importance of unresolved small surface features for computation of YORP, and demonstrated that structures as small as one tenth of the radius of an asteroid can crucially change the amount of its YORP acceleration. But the importance of much smaller structures with sizes below one meter for rotational dynamics of big asteroids still hasn't been discussed. However, at these scales the physics of YORP can drastically change.

For coarse-grained shape models, heat conductivity under each element of the surface appears to be effectively 1-dimensional (Breiter et al., 2010a; Golubov & Krugly 2010). Then different surface elements don't exchange heat with each other, the acceleration of the asteroid doesn't depend on the thermal model, and the Rubincam approximation (which assumes instant re-emission of light by the surface) is precise for computing the YORP acceleration et al. In contrast, when decimeter-sized asteroids are considered, the heat exchange between different surface elements switches on, and the Rubincam approximation breaks (Breiter et al., 2010b).

But heat conductivity in the asteroid's body is important not only for decimeter-sized asteroids, but also for big asteroids that have decimeter-sized stones on the surface. It occurs, that light re-emitted by these stones can create a force, dragging the surface of the asteroid in tangential direction. Even though each small stone experiences recoil forces only normal to its surface, when we view the surface globally on the scales of meters or tens of meters, non-compensated forces acting on different sides of stones add up to create a force, which has a component parallel to the smoothed 'global' surface. We use the term "tangential YORP" (or "T-YORP") for the effect of these tangential forces on the rotation of the asteroid, to distinguish it from the "normal YORP" (or "N-YORP"), created by forces normal to the 'global' surface.

Two distance scales are particularly important for our consideration. One of them is the heat conductivity length,

$$L_{cond} = \frac{\kappa}{\sqrt[4]{(1-A)^3 \Phi^3 \varepsilon \sigma}}. \tag{1}$$

Here $\kappa$ is the heat conductivity, $A$ is the albedo of the surface, $\varepsilon$ is its thermal emissivity, $\Phi$ is the solar energy flux, $\sigma$ is Stefan–Boltzmann's constant. Equation (1) is obtained by equating the incoming energy flux to the heat flux, which corresponds to the temperature difference equal to the temperature at the subsolar point over the distance $L_{cond}$. Heat conductivity through structures much bigger than $L_{cond}$ comprises a negligible part of the incident heat.

The second important distance scale is the thermal wavelength,

$$L_{wave} = \sqrt{\frac{\kappa}{C\rho\omega}}. \tag{2}$$

Here $C$ is the heat capacity of the soil, $\rho$ is its density, and $\omega$ is the angular velocity of the asteroid.

The ratio of these two scales defines the thermal parameter,

$$\theta = \frac{L_{cond}}{L_{wave}} = \frac{(C\rho\kappa\omega)^{1/2}}{((1-A)\Phi)^{3/4}(\varepsilon\sigma)^{1/4}}. \tag{3}$$

The case $\theta \gg 1$ implies that the temperature of the surface stays almost the same in the course of the asteroid's rotation. It can occur if the heat conductivity is high or if the asteroid's rotation is fast. In the opposite case $\theta \ll 1$ the temperature of the surface at each instant of time is almost in equilibrium with the incident radiation, and is defined by the Stefan–Boltzmann law.

## 2. The simplest model: sphere with meridional walls

The simplest model in which the tangential YORP can be observed is presented in the upper panel of Figure 1. It is somewhat similar to the model proposed by Rubincam (2000), though perfectly symmetric. A spherical asteroid has two vertical stone walls standing across its equator in the meridional direction. The equatorial plane of the asteroid is assumed to coincide with its orbital plane. The orbit is assumed to be circular. To account for self-illumination effects we assume mirror reflection of the sunlight from the surface, so that each incident light ray is instantly reflected by the surface with the reflection angle equal to the incidence angle. In reality the angular distribution of the light scattered by the surface is more complicated, and also part of the incident light is re-emitted in infrared, with a different indicatrix and some time delay.

To calculate the mean pressure acting on the surface of a wall we have to solve the heat conductivity equation in the wall, to find the light emitted by its surfaces and the pressure it exerts, and to average this pressure over the rotation period. We assume the heat conductivity in the wall to be one-dimensional that is justified only if the height of the wall $h$ and the length $l$ are much greater than its thickness $d$. We introduce the coordinate $x$ ranging through the wall from 0 (the east side of the wall) to $d$ (the west side), as it is shown in the middle panel of Figure 1. In this article we *define* east and west so that the Sun rises in the east and sets in the west, regardless of the sense of rotation of the asteroid with respect to its orbit or to the Earth.

We normalize the coordinate $x$ by the heat conductivity length $L_{\text{cond}}$, temperature $T$ by the equilibrium temperature of the subsolar point, and use the rotation phase $\varphi$ instead of the time $t$, thus getting a new set of variables,

$$\zeta = \frac{x}{L_{\text{cond}}}, \ \tau = \sqrt[4]{\frac{\varepsilon\sigma}{(1-A)\Phi}} T, \ \varphi = t\omega. \tag{4}$$

Then the heat conductivity equation in these dimensionless variables looks like

$$\frac{\partial \tau}{\partial \varphi} = \frac{1}{\theta^2}\frac{\partial^2 \tau}{\partial \zeta^2}. \tag{5}$$

The boundary conditions on the eastern and western sides of the surface are

$$\left.\frac{\partial \tau}{\partial \zeta}\right|_{\zeta=0} = 2\sin\varphi H(\sin\varphi)H(-\cos\varphi) - \tau^4, \tag{6}$$

$$\left.\frac{\partial \tau}{\partial \zeta}\right|_{\zeta=1} = -2\sin\varphi H(-\sin\varphi)H(-\cos\varphi) - \tau^4, \tag{7}$$

Here $H$ is Heaviside step-function. The first Heaviside function accounts for shadowing of the solar light by the wall, the second one accounts for shadowing by the asteroid's surface at night. The factor 2 stands for self-illumination of the wall by the asteroid's surface reflecting the solar light in our simplified model of mirror reflection.

Solving heat conductivity Equation (5) with boundary conditions Equations (6)-(7), we get the temperature distribution inside the wall. This temperature distribution in different instants of time for the case $d=L_{cond}=L_{wave}$ is presented in the upper left panel of Figure 2. The time is given in "asteroid hours", so that sunrise is at 6 a.m., midday at 12, and sunset at 18. Naturally, the temperature of the east side of the wall is higher in the morning; the temperature of the west side is higher in the afternoon. But we can notice, that the temperature of the west face reaches higher values than the temperature of the east face.

The temperature $\tau$ leads to light emission from the surface, which creates the pressure $p=\Phi\Pi/c$, with $\Pi=2/3\theta^4$ being the dimensionless pressure. $\Pi$ at the two sides of the wall as a function of time is presented in the upper right panel of Figure 2. The difference between these two pressures $\Delta\Pi$ is plotted in green. For the selected parameters the mean dimensionless pressure over the period appears to be positive. Its mean value is plotted with a red line, which is hardly distinguishable from the $x$-axis.

This slight net pressure difference $<\Delta\Pi>$ creates the tangential YORP. If $<\Delta\Pi>$ is positive, the wall accelerates the rotation of the asteroid. If $<\Delta\Pi>$ is negative, the asteroid is decelerated. $<\Delta\Pi>$ is presented in the lower left panel of Figure 2 as a function of $\lg\theta$ for different values of $\lg(d/L_{cond})$, and in the lower right panel of Figure 2 $<\Delta\Pi>$ is color-coded as a function of both $\lg\theta$ and $\lg(d/L_{cond})$. We can see that the maximal amount of the tangential YORP $<\Delta\Pi>\approx 0.014$ is attained for $d/L_{cond}\sim 1$ and $\theta\sim 1$. When one moves away from the neighbourhood of this point in either direction, $\langle\Delta\Pi\rangle$ decreases.

To compare relative amounts of tangential and normal components of YORP, it is convenient to normalize the YORP torque over a specific torque $T_0=\Phi r^3/c$, where $r$ is the equivalent radius of the asteroid (the radius of the sphere of the same volume), $c$ is the speed of light, $\Phi$ is the average solar radiation flux illuminating the asteroid. The dimensionless YORP torque $\tau_z=T_z/T_0$ appears to be a convenient measure of the strength of the YORP effect, with $T_z$ being the YORP torque with respect to the rotation axis $z$ of the asteroid. In the Rubincam approximation only highly asymmetric shapes can produce high values of $\tau_z$, and $\tau_z$ can be considered as a rate of non-symmetry of the asteroid's shape, independent of its size and orbit. The highest possible values $\tau_z\approx 1$ can be reached only for hypothetical bodies, whose surface re-emit all the incident light in the west direction.

The dimensionless torque $\tau_z$ can be estimated from the observed angular acceleration $\dot\omega$ via formula

$$\tau_z \approx \frac{\pi(A_x^2 + A_y^2)\rho c\dot\omega}{15\Phi}, \qquad (8)$$

where $A_x$ and $A_y$ are the longest and the intermediate axes of the asteroid's body, $\rho$ is its density, and $\Phi$ is solar radiation flux at the distance of the asteroid orbit's major semi-axis. Thus we can estimate $\tau_z=0.008$ for 1620 Geographos (Ďurech et al. 2008) and $\tau_z=0.002$ for 54509 YORP (Lowry et al. 2007).

To estimate the tangential YORP acceleration, let's imagine a spherical asteroid with parallel walls going from north to south, with height of each wall being $h$ and the distance between two neighbouring walls being $a$ (middle panel of Figure 1). Two walls going along the same meridian at different latitudes absorb the same amount of solar light per unit surface of the wall. Thus boundary conditions of heat conductivity equation are the same, so are their solutions. If we assume that the pressure difference between the west and the east side of each wall is $\langle p\rangle=\Phi\langle\Pi\rangle/c$ and integrate the YORP torque over the asteroid surface, we get for dimensionless YORP torque $\tau_z=\pi^2\langle\Pi\rangle h/a$. Let us assume $h/a=1/3$. It's questionable whether an asteroid can have such rough surface, but at least this is enough to avoid strong shadowing of walls and to use values for Π from the two bottom panels of Figure 2 as an estimate. Thus we get $\tau_z \approx 3\Pi$, implying that $\tau_z$ can reach values up to 0.04 if the parameters have appropriate values. This appears to be even more than the observed amounts of the YORP acceleration.

## 3. Discussion.

If instead of our idealized walls we consider stones of realistic shapes, the amount of T-YORP can appear to be essentially less than in our estimate, but the exact magnitude of the effect requires further investigation. Still, we already can extract much important information about general features of T-YORP from our simplified model.

We can see in Figure 2 that the effect is significant only for $d/L_{cond} \sim 1$ and $\theta \sim 1$. It is easy to see the reason why there should be no T-YORP if any of these two conditions violates:

- If the asteroid rotates too fast ($\theta \gg 1$), temperature of each face of the wall stays nearly constant, and is then determined by the balance between the overall daily illumination and the relaxed time-independent heat flux. As the former is the same for the east and the west face, the temperature also appears to be the same, with no T-YORP torque.

- If the asteroid rotates too slow ($\theta \ll 1$), the material demonstrates no thermal inertia, with the temperature at each instant of time being determined by the balance between the instantaneous illumination and the relaxed heat flux. Then temperature of the east face in the morning is the same as the temperature of the west side in the evening and vice versa, and the net effect also vanishes.

- If the wall is very thick ($d/L_{cond} \gg 1$), the heat conductivity in the wall is negligible, and each face of the wall emits as much energy as it has absorbed. Thus the two faces have the same net emission, which exerts the same net pressure, and compensate the torques of each other.

- If the wall is too thin ($d/L_{cond} \ll 1$), the heat conductivity in the wall is sufficient to make temperatures of two faces equal at every moment, with the T-YORP torque at every moment being zero.

In contrast, in the intermediate case, when $d/L_{cond} \sim 1$ and $\theta \sim 1$, resonant-like effects can occur, with heat wave from one side of the wall transporting a significant amount of energy to the other side, and creating T-YORP torque. From Figure 1 we can also see that for a given $\theta$ the effect is the strongest when $d/L_{cond} \approx 1/\theta$, or $d \approx L_{wave}$, that is qualitatively consistent with picturing T-YORP as a resonant phenomenon.

With Table 1 we can see for which surface structures both conditions $d/L_{cond} \sim 1$ and $\theta \sim 1$ are best satisfied. Here we estimate $L_{cond}$, $L_{wave}$, and $\theta$ in different cases: for regolith, basalt and iron-reach material (Farinella et al. 1998), for near-Earth ($a = 1$ AU) and main belt ($a = 3$ AU) asteroids, for fast ($T=3$h) and slow ($T=30$h) rotators. Comparing values from Table 1 with Figure 1 we see that regolith could produce strong T-YORP, but only if were reach in structures several millimeters in diameter. It is questionable whether structures of this size are abundant on regolith, as well as whether our approximation treating surface as a continuum will break at such small scales. In any case, it's important to realize that even such small scales can produce a strong YORP acceleration. Basaltic stones with size about a decimeter can create a strong T-YORP acceleration of near-Earth slow rotators. Iron-reach stones also create the strongest T-YORP acceleration in the case near-Earth, with size of stones around several decimetres, but the value of the acceleration is much smaller than in the previous two cases. The biggest possible acceleration can be expected for thermal conductivities intermediate between basalt and regolith, which could correspond to a strongly eroded stone.

In most cases T-YORP leads to acceleration of asteroids, decreasing rotation periods of both prograde and retrograde rotators. Though for some parameters we also get negative acceleration (see Figure 1), it is unclear whether it will persist for realistic 3-dimensional models of stones.

An interesting behaviour can occur, if an asteroid has a negative N-YORP acceleration. This torque is independent of the asteroid's circular velocity $\omega$ (Scheeres 2007; Golubov & Krugly 2010; Breiter et al. 2010a), while T-YORP acceleration depends on circular velocity via $\theta$, that is proportional to $\omega^{1/2}$. Dependence of T-YORP pressure for a single wall on $\theta$ is plotted in the lower left panel of Figure 2. To plot the total YORP acceleration we have to pick in the figure the line corresponding to sizes of stones, which we have at the asteroid, rescale it vertically proportionally to the surface density of stones, and shift down to account for N-YORP contribution. If the negative contribution of N-YORP and the positive contribution of T-YORP are of the same order of magnitude, we can have the dependence of YORP on $\theta$ intersecting the *x*-axis in two points. The second point of intersection is stable (in contrast to the first one): if $\theta$ (and thus $\omega$) increases, YORP becomes negative, and increases $\omega$ to restore the equilibrium. Therefore, we must conclude that if our understanding of the concept of N-YORP is correct, we can expect to observe a significant fraction of asteroids with precisely zero YORP acceleration, whose rotation state evolved in such a way to be locked in an equilibrium state.

Asteroid 25143 Itokawa could in principle be one of these cases. From the Hayabusa space mission its shape is known with unprecedented precision (Gaskell et al., 2008). When using these shape models to predict N-YORP, small negative amounts are obtained (Scheeres et al., 2007; Scheeres & Gaskell 2008), with $\tau_z$ ranging from $-13 \times 10^{-4}$ to $-2 \times 10^{-4}$ depending on the used model. But the observations reveal no YORP deceleration of the asteroid (Durech et al., 2008), $\tau_z = (0.4\pm4) \times 10^{-4}$. The discrepancy of ~0.001 seems quite possible for T-YORP, whose contribution according to our idealized model can be as big as 0.04, while the surface of Itokawa is known to be quite rocky. It must be mentioned, that more 'conservative' explanations of the discrepancy are also possible, such as decelerating influence of close encounter with the Earth in 2004 (Durech et al., 2008) and offset of center of mass of the asteroid with respect to its geometric center (Scheeres et al., 2008).

When we think about T-YORP globally, on the scale of tens of meters, we can equivalently say that it is produced by non-symmetry of the re-emission indicatrix with respect to the normal vector of the smoothed global surface. Due to heat conductivity processes in stones, the indicatrix can be inclined westwards, thus creating a force that has an eastward component. The pressure forces creating T-YORP, N-YORP, and the total YORP are sketched in the bottom panel of Figure 1. The T-YORP force is about 2 orders of magnitude smaller than the N-YORP force acting along the normal of the surface. But N-YORP forces act more or less in the direction of the asteroid's center, thus having smaller lever arms. Moreover, the N-YORP torques produced by different parts of the surface have different signs, and largely compensate each other, while all T-YORP torques can have the same sign, add up, and overcome N-YORP at the end. The N-YORP torque is produced by the slight non-ellipticity of an asteroid. In contrast, even a spherical asteroid can experience T-YORP.

Westward inclination of the re-emission indicatrix is not the only consequence of heat conductivity processes in stones. The indicatrix must also be inclined towards equator, as sides of stones that are closer to equator are better illuminated. Also the mean normal pressure can be altered by heat conductivity processes in stones. But we expect these effects to be less important than the one we have discussed, as their torques created by different parts of the surface also largely compensate each other. The situation here is similar to the case of anisotropic re-radiation indicatrix considered by Breiter & Vokrouhlický (2011), who demonstrated that the dependence of the indicatrix on the direction of the incident radiation doesn't change the N-YORP acceleration much. In contrast to their results, in this paper we have argued that not only the direction of the incident radiation, but also the rotation of the asteroid can break symmetry of its indicatrix, and the latter asymmetry can significantly alter the YORP acceleration.

**Acknowledgements.** OG is very grateful to his friends A. Tkachuk and M. Tkachuk, and his teacher C. Dullemond, without whose erudition in numerical methods and readiness to help he could hardly create the programs, whose results are presented in this article. Authors are very grateful to D. Scheeres for reading and commenting on the manuscript.

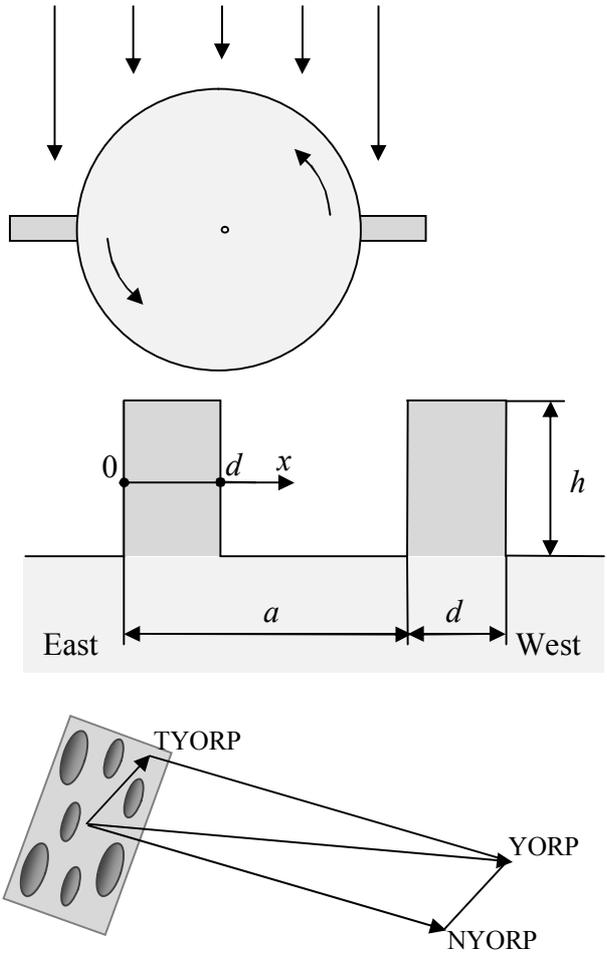

Figure 1. *Top:* The simplest model of an asteroid subject to the effect. It is a spherical asteroid with two vertical walls standing near its equator in the meridional direction. The view from the north pole. *Middle:* A simplified physical model to calculate T-YORP: asteroid with meridian walls. The figure presents the asteroid's cross-section by a vertical plane in the direction from east to west. Vertical stone walls of thickness $d$ and height $h$ cross the surface in meridian direction at the distance $a$ from each other. The surface is covered with regolith. *Bottom:* A sketch of the normal and tangential components of the pressure forces acting on the surface to produce N- and T- contributions to the YORP effect.

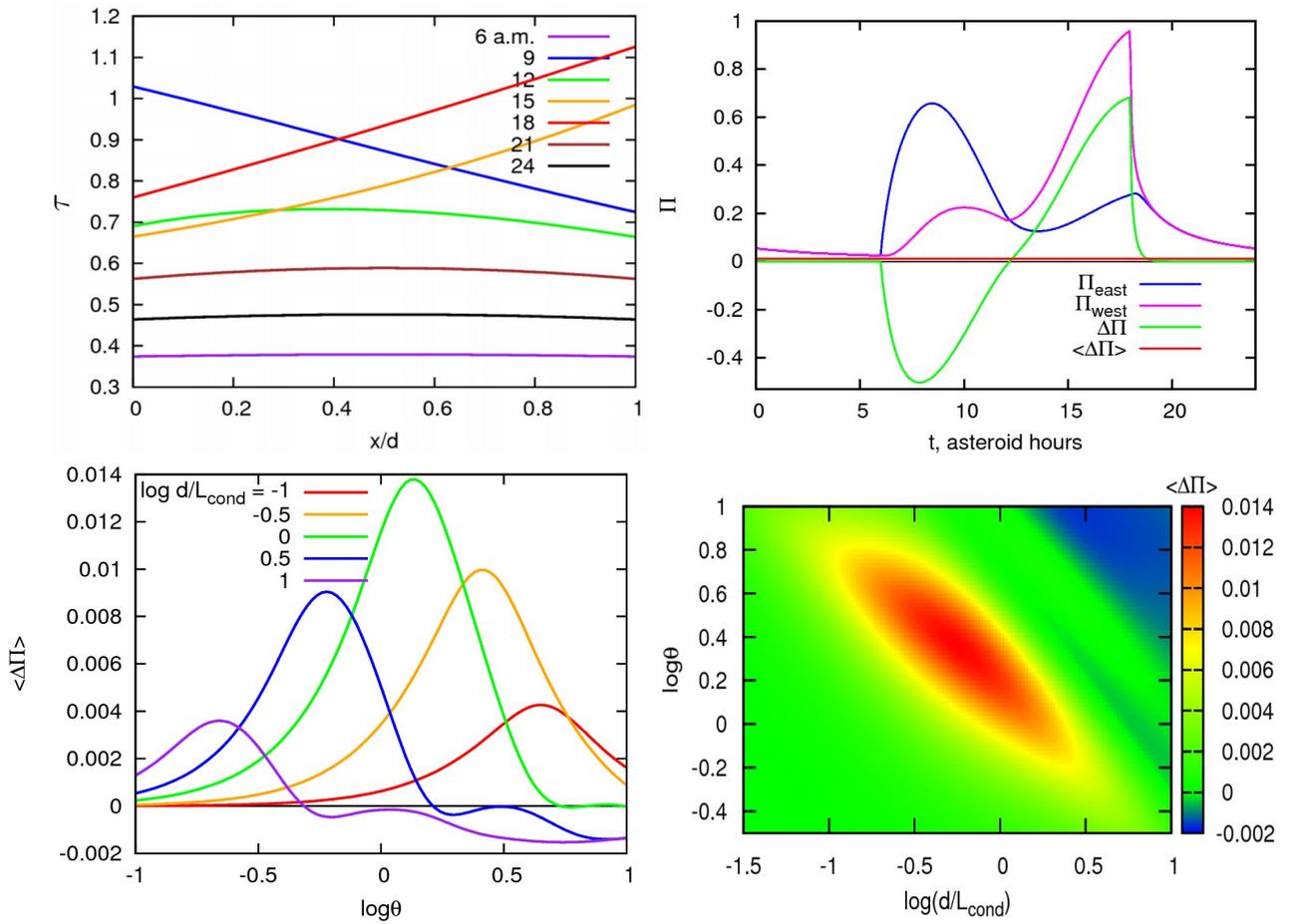

Figure 2. Emergence of T-YORP in the simplified model of 1-dimensional wall. *Upper left panel:* Temperature distribution in the wall with $d = L_{\text{cond}} = L_{\text{wave}}$ at different instants of time. The left edge of the plot is the east face of the wall, the right edge is the west (prograde rotation). The time is given in "asteroid hours". *Upper right panel:* Pressures acting on the west and on the east sides of the asteroid as functions of time are plotted in blue and magenta. Their difference is plotted in green. The difference averaged over the rotational period is plotted with a red horizontal line only slightly above the *x*-axis. *Lower left panel:* Mean pressure difference acting on the wall plotted against the thermal parameter for several different thicknesses of the wall. *Lower right panel:* Mean pressure difference acting on the wall color-coded as a function of the thicknesses of the wall and the thermal parameter.

Table 1. Estimates of the heat conductivity length $L_{cond}$, the thermal wavelength $L_{wave}$, and the thermal parameter $\theta$ in different cases. Properties of regolith, basalt and iron-reach material are taken from Farinella et al. (1998). Other parameters are assumed to be $A=0.3$, $\varepsilon=0.7$, $\Phi=1400 W \cdot m^{-2} AU^2/a^2$, and $\sigma=5.67 \cdot 10^{-8} W \cdot m^{-2} K^{-4}$.

| Material | $a$, AU | $P$, h | $L_{cond}$, cm | $L_{wave}$, cm | $\theta$ |
|---|---|---|---|---|---|
| Regolith | 1 | 3 | 0.06 | 0.15 | 0.4 |
| $\kappa= 0.0015$ W m$^{-1}$K$^{-1}$ | 1 | 30 | 0.06 | 0.5 | 0.12 |
| $C=680$ J kg$^{-1}$K$^{-1}$ | 3 | 3 | 0.3 | 0.15 | 2 |
| $\rho=1500$ kg m$^{-3}$ | 3 | 30 | 0.3 | 0.5 | 0.6 |
| Basalt | 1 | 3 | 110 | 4 | 25 |
| $\kappa=2.65$ W m$^{-1}$ K$^{-1}$ | 1 | 30 | 110 | 14 | 8 |
| $C=680$ J kg$^{-1}$K$^{-1}$ | 3 | 3 | 560 | 4 | 130 |
| $\rho=3500$ kg m$^{-3}$ | 3 | 30 | 560 | 14 | 40 |
| Iron-rich | 1 | 3 | 1600 | 13 | 120 |
| $\kappa=40$ W m$^{-1}$ K$^{-1}$ | 1 | 30 | 1600 | 40 | 40 |
| $C=500$ J kg$^{-1}$K$^{-1}$ | 3 | 3 | 8400 | 13 | 640 |
| $\rho=8000$ kg m$^{-3}$ | 3 | 30 | 8400 | 40 | 200 |